\documentclass[a4paper,11pt]{article}
\usepackage{pos}
\usepackage{graphicx}
\usepackage{psfrag}
\usepackage{color}
\usepackage{hyperref}
\hypersetup{colorlinks=true,linkcolor=purple,anchorcolor=blue,citecolor=blue, filecolor=blue,urlcolor=red,bookmarksnumbered=true,
	pdfview=FitB
}
\usepackage{subcaption}
\usepackage{xcolor}
\usepackage{ulem}
\usepackage{csquotes}
\colorlet{purple1}{blue!70!red}
\colorlet{darkred}{red!50!black}
\usepackage{graphicx}
\usepackage{psfrag}
\usepackage{color}
\usepackage{comment}
\usepackage{amsmath}
\usepackage{epstopdf}
\usepackage{epsf}

\title{Proton gluonic distributions in a light front spectator model}

\author*[a,b]{Poonam~Choudhary}
\author[a]{Dipankar~Chakrabarti}
\author[a]{Bheemsehan~Gurjar}
\author[c]{Tanmay~Maji}
\author[d,e]{Chandan~Mondal}
\author[b]{Asmita~Mukherjee}

\affiliation[a]{Department of Physics, Indian Institute of Technology Kanpur \\
  Kanpur-208016, India}

\affiliation[b]{Department of Physics, Indian Institute of Technology Bombay \\
 Powai, Mumbai 400076, India}

 \affiliation[c]{Department of Physics, National Institute of Technology Kurukshetra \\ Kurukshetra-136119, India }

 \affiliation[d]{Institute of Modern Physics, Chinese Academy of Sciences \\
 Lanzhou 730000, China}

 \affiliation[e]{School of Nuclear Science and Technology, University of Chinese Academy of Sciences, \\
 Beijing 100049, China}

\emailAdd{poonam.hep.ph@gmail.com}
\emailAdd{dipankar@iitk.ac.in} 
\emailAdd{gbheem@iitk.ac.in} 
\emailAdd{tanmayphy@nitkkr.ac.in}
\emailAdd{mondal@impcas.ac.cn}
\emailAdd{asmita@phy.iitb.ac.in} 

\abstract{ { We investigate the leading-twist gluonic distribution functions within the proton by formulating a light-front spectator model. The proton light-front wave functions (LFWFs) are adopted from the soft-wall AdS/QCD predictions, and the model parameters are determined using the gluon unpolarized parton distribution function from the NNPDF3.0nlo dataset. Furthermore, we demonstrated the model calculations for gluon transverse momentum distributions (TMDs) and generalized parton distributions (GPDs). We also predicted the gluon spin and orbital angular momentum (OAM) contributions to the total proton spin.}}

\FullConference{%
  31st International Workshop on Deep Inelastic Scattering (DIS2024)\\
   8–12 April 2024 \\
  Grenoble, France
}

\begin{document}

\maketitle

\section{Introduction}\label{sec:Intro}
{TMDs depict the proton's three-dimensional momentum space landscape and are crucial for understanding the dynamics of quarks and gluons within the proton~\cite{Meissner:2007rx}. Over the past two decades, significant progress has been made in studying quark TMDs (for a recent review on quark TMDs please see~\cite{Boussarie:2023izj} and references therein), while gluon TMDs in the proton remain less explored~\cite{Bacchetta:2020vty,Lu:2016vqu,Yu:2024mxo}. Whereas, on the other hand, GPDs also provide a three-dimensional interpretation of parton distributions and help in understanding the contributions of gluons to the nucleon's mass, spin, and mechanical properties~\cite{Diehl:2003ny}. TMDs depend on the longitudinal momentum fraction $x$ and the transverse momentum $\bfp$, and can be accessed through Semi-Inclusive Deep Inelastic Scattering (SIDIS) and Drell-Yan (DY) processes, while GPDs are functions of the longitudinal momentum fraction $x$, momentum transfer $\Delta$, and the skewness parameter $\xi$, and can be studied through Deeply Virtual Compton Scattering (DVCS). Like TMDs, quark GPDs have been extensively studied over the years~\cite{Guidal:2013rya}. In contrast, there are relatively few studies on gluon GPDs available in the literature~\cite{Tan:2023kbl,Lin:2023ezw,Kriesten:2021sqc}.  
} In this contribution, we focus on the three-dimensional gluon distribution within the proton, exploring both TMDs \cite{Chakrabarti:2023djs} and GPDs \cite{Chakrabarti:2024hwx}. 
\vspace{-0.35cm}
\section{Gluon spectator model }\label{sec:model}
In this section, we present a light-front spectator model ~\cite{Chakrabarti:2023djs} for a proton, wherein the gluon is regarded as an active parton while the remnant known as the spin-1/2 spectator system. 
The model utilizes on the light-cone formalism in which the proton momentum is given as  $ P=(P^+,\frac{M^2}{P^+},\textbf{0}_\perp)$ where the transverse momentum of the proton is considered zero.  
The active parton and spectator momentum are given by $ p=(x P^+, \frac{p^2+\textbf{p}_\perp^2}{x P^+}, \textbf{p}_\perp)$ and $ P_X=((1-x) P^+, P^-_X, -\textbf{p}_\perp)$ respectively. The variable $x=p^{+}/P^{+}$ and ${\bf p}_{\perp}$ represent the fraction of longitudinal momentum and transverse momentum respectively for the active gluon. 
The proton state can therefore be expressed as a two-particle Fock-state expansion ~\cite{Brodsky:2000ii} with proton helicity $J_{z}=\pm\frac{1}{2}$, 
 \begin{eqnarray}\label{state}\nonumber
		|P;\uparrow(\downarrow)\rangle
		= \int \frac{\mathrm{d}^2 \bfp \mathrm{d} x}{16 \pi^3 \sqrt{x(1-x)}}\times \Bigg[\psi_{+1+\frac{1}{2}}^{\uparrow(\downarrow)}\left(x, \bfp\right)\left|+1,+\frac{1}{2} ; x P^{+}, \bfp\right\rangle+\psi_{+1-\frac{1}{2}}^{\uparrow(\downarrow)}\left(x, \bfp \right)\left|+1,-\frac{1}{2} ; x P^{+}, \bfp \right\rangle\\ 
		+\psi_{-1+\frac{1}{2}}^{\uparrow(\downarrow)}\left(x, \bfp \right)\left|-1,+\frac{1}{2} ; x P^{+}, \bfp \right\rangle+\psi_{-1-\frac{1}{2}}^{\uparrow(\downarrow)}\left(x, \bfp\right)\left|-1,-\frac{1}{2} ; x P^{+}, \bfp\right\rangle\bigg],
	\end{eqnarray}
	where $\psi_{\lambda_{g}\lambda_{X}}^{\uparrow(\downarrow)}(x,\bfp)$ are 
 probability amplitudes of corresponding two-particle Fock state $|\lambda_{g},\lambda_{X};xP^{+},\bfp \rangle$, and 
$\lambda_{g}$, $\lambda_{X}$ are the helicities of active gluon and spectator respectively. The complete LFWFs expression for a proton with $J_{z}= \pm 1/2$ are given by~\cite{Chakrabarti:2023djs,Chakrabarti:2024hwx}: 
 \begin{eqnarray} \label{LFWFsuparrow}   \nonumber
		\psi_{+1+\frac{1}{2}}^{\uparrow}\left(x,\bfp\right)&=&-\sqrt{2}\frac{(-p^{1}_{\perp}+ip^{2}_{\perp})}{x(1-x)}\varphi(x,\bfp^2), \hspace{1cm} \psi_{+1+\frac{1}{2}}^{\downarrow}\left(x, \bfp\right)= 0, \\ \nonumber
		\psi_{+1-\frac{1}{2}}^{\uparrow}\left(x, \bfp\right)&=&-\sqrt{2}\bigg( M-\frac{M_{X}}{(1-x)} \bigg) \varphi(x,\bfp^2), \hspace{0.7cm} \psi_{+1-\frac{1}{2}}^{\downarrow}\left(x,\bfp\right)=-\sqrt{2}\frac{(-p^{1}_{\perp}+ip^{2}_{\perp})}{x}\varphi(x,\bfp^2),  \\ \nonumber
		\psi_{-1+\frac{1}{2}}^{\uparrow}\left(x, \bfp\right)&=&-\sqrt{2}\frac{(p^{1}_{\perp}+ip^{2}_{\perp})}{x}\varphi(x,\bfp^2), \hspace{0.8cm} \psi_{-1+\frac{1}{2}}^{\downarrow}\left(x, \bfp\right)=-\sqrt{2}\bigg( M-\frac{M_{X}}{(1-x)} \bigg) \varphi(x,\bfp^2), \\
		\psi_{-1-\frac{1}{2}}^{\uparrow}\left(x, \bfp\right)&=&0, \hspace{4.5cm} \psi_{-1-\frac{1}{2}}^{\downarrow}\left(x, \bfp \right)= -\sqrt{2}\frac{(p^{1}_{\perp}+ip^{2}_{\perp})}{x(1-x)}\varphi(x,\bfp^2). 
	\end{eqnarray}
where the modified form of two particle soft-wall AdS/QCD wave function, $\varphi(x,\mathbf{p}_{\perp}^2)$ is given as,
\be\label{AdSphi}
	\varphi(x,\bfp^2)=N_{g}\frac{4\pi}{\kappa}\sqrt{\frac{\log[1/(1-x)]}{x}}x^{b}(1-x)^{a}\exp{\bigg[-\frac{\log[1/(1-x)]}{2\kappa^{2}x^2}\bfp^{2}\bigg]}.
	\ee	
\begin{figure}
\includegraphics[scale=0.5]{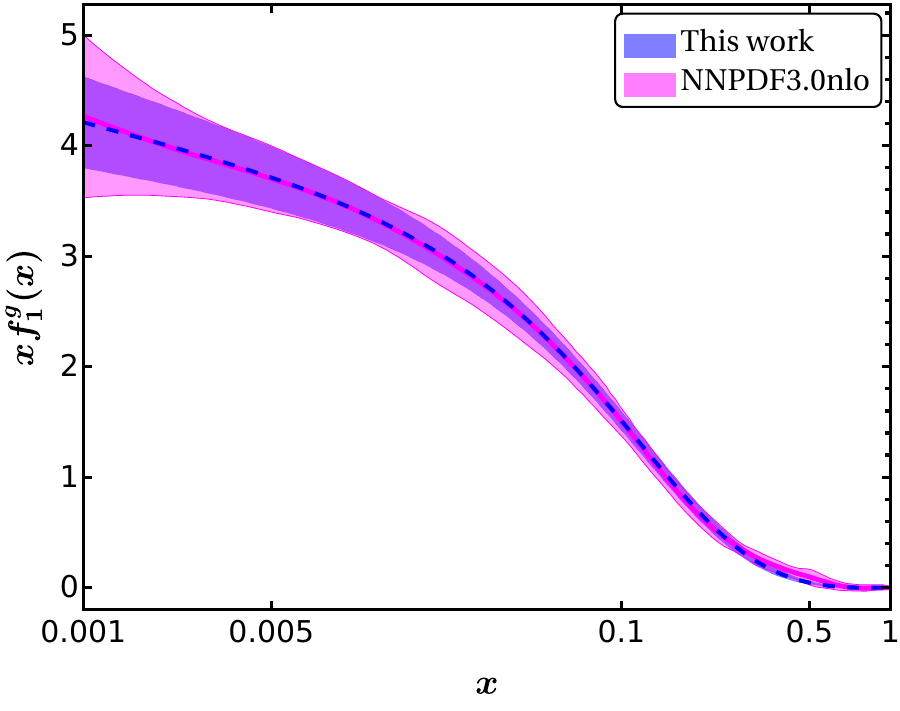}
	\hspace{0.3cm}
 \includegraphics[scale=0.52]{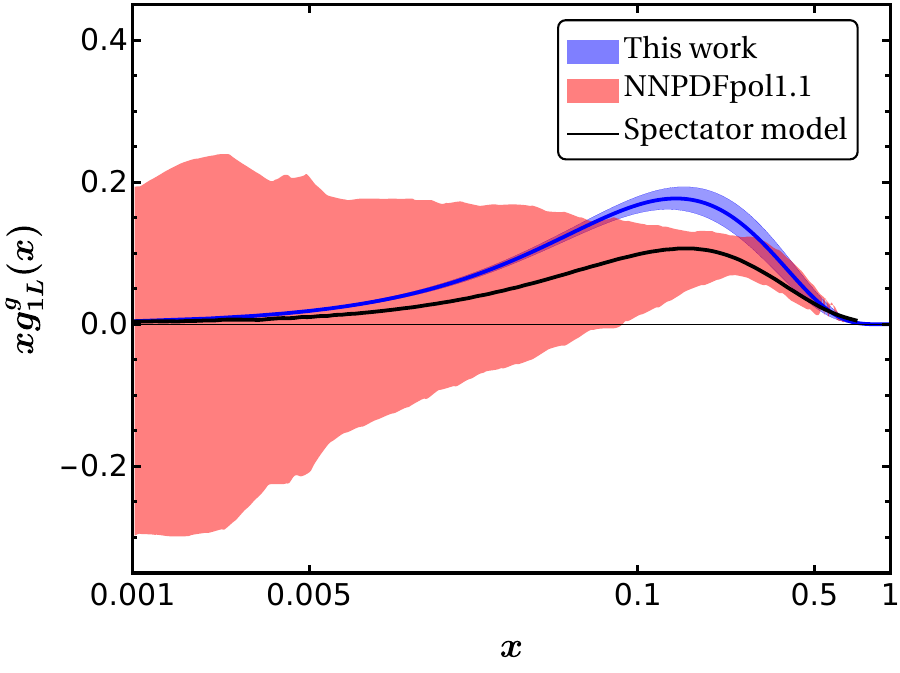}
 \caption{Left plot shows fitting of the unpolarized gluon PDF from the model with the NNPDF3.0nlo dataset~\cite{NNPDF:2017mvq} while in the right plot, the results of NNPDFpol1.1~\cite{Nocera:2014gqa} and the spectator model ~\cite{Bacchetta:2020vty} are compared with our model predictions of gluon helicity PDF. The plots are taken from \cite{Chakrabarti:2023djs}.}
	\label{fig:helicityasymmetry}
\end{figure}
 where the model information is encoded in $a$ and $b$ parameters, and the soft-wall AdS/QCD scale parameter is taken as $\kappa=0.4$ GeV \cite{Chakrabarti:2013gra}. The model parameters $a$, $b$, and the normalization constant $N_{g}$ are fixed at the scale $Q_{0}=2$ GeV by fitting the model unpolarized PDF with NNPDF3.0 data \cite{NNPDF:2017mvq}. 
 Using the proton LFWFs and transverse momentum integrated correlation function, we obtained the unpolarized and helicity gluon PDFs and found that at the end points, i.e., $x\rightarrow 0,1$, the ratio of gluon helicity PDF to unpolarized PDF satisfy the pQCD constraints. We also predicted the gluon spin contribution, $\Delta G$, to the proton spin ~\cite{Chakrabarti:2023djs}. 
 
\vspace{-0.35cm}
\section{Transverse Momentum Distributions (TMDs)} 
In this section, we present the leading-twist T-even gluon TMDs derived by employing the unintegrated gluon correlation function within the light-front formalism. The gluon TMD correlator in SIDIS process is given as,
\begin{eqnarray} \label{eq:correlator}
		\Phi^{g[ij]}(x,\bfp;S)
		=\frac{1}{xP^+}\int\frac{d\xi^-}{2\pi}\,\frac{d^2 \mathbf{\xi}{_\perp}}{(2\pi)^2}\,e^{i k\cdot \xi}\,
		\big<P;S\big|\,
		F^{+j}_a(0) \ \,\mathcal{W}_{+\infty,ab}(0;\xi)\,
		F^{+i}_b(\xi)\,\big|P;S\big>\,
		\Big|_{\xi^+=0^+} \,,
	\end{eqnarray}
Where $|P;S\rangle$ represents the proton state with momentum $P$ and spin $S$, $F^{\mu\nu}_{a}$ is the gluon field operator, and $\mathcal{W}_{+\infty,ab}$ is the future-pointing Wilson line that confirms the gauge invariance of the correlator. Depending on the possible gluon and nucleon polarizations, there exists eight gluon TMDs at leading-twist. Four of them ($f_{1}^{g},~g_{1L}^{g},~g_{1T}^{g}$ and $h_{1}^{\perp g}$) are T-even. 
Furthermore, by employing the overlap representation of LFWFs, we obtained all four gluon T-even TMDs in our previous work~\cite{Chakrabarti:2023djs}. We found that the three-dimensional distribution of all four T-even leading twist gluon TMDs with longitudinal momentum fraction $x$ and transverse momentum $\bfp$ exhibits a positive peak at small $x$, but falls sharply for higher $\bfp$ and large $x$. We also verified the positivity bound conditions and various other model dependent and model independent relations among the T-even gluon TMDs ~\cite{Chakrabarti:2023djs}.
\vspace{-0.35cm}
\section{Generalized Parton Distribution (GPDs) at non-zero skewness}
 The off-forward matrix elements of the bilocal currents can be parameterized in terms of eight leading twist gluon GPDs in the light cone gauge where $A^{+}=0$ ~\cite{Diehl:2003ny}. These GPDs are further divided into chiral-even and chiral-odd categories depending on the helicity states of the initial and final state of the gluon. The chiral-even (helicity non-flip) GPDs are parameterized as following: 
\begin{align}
 \frac{1}{P^+} \int \frac{d z^-}{2\pi}\, e^{ix P^+ z^-}
  \langle p',\lambda'|\, 
     F^{+i}(-{\textstyle\frac{z}{2}})\, 
     F^{+i}({\textstyle\frac{z}{2}})\, 
  \,|p,\lambda \rangle \Big{|}_{\substack{z^+=0\\\mathbf{z}_{T}=0}}= \frac{1}{2P^+} \bar{u}(p',\lambda')
  \left[
  H^g\, \gamma^+ +
  E^g\, \frac{i \sigma^{+\alpha} \Delta_\alpha}{2M} 
  \right] u(p,\lambda) ,\\
   - \frac{i}{P^+} \int \frac{d z^-}{2\pi}\, e^{ix P^+ z^-}
  \langle p',\lambda'|\, 
     F^{+i}(-{\textstyle\frac{z}{2}})\, 
          \tilde{F}^{+i}({\textstyle\frac{z}{2}})\, 
  \,|p,\lambda \rangle \Big{|}_{\substack{z^+=0\\\mathbf{z}_{T}=0}}
= \frac{1}{2P^+} \bar{u}(p',\lambda') \left[
  \tilde{H}^g\, \gamma^+ \gamma_5 +
  \tilde{E}^g\, \frac{\gamma_5 \Delta^+}{2M}\, \, 
  \right] u(p,\lambda) ,
\end{align}
where $\tilde{F}^{\alpha\beta}=\frac{1}{2}\epsilon^{\alpha\beta\gamma\delta}F_{\gamma\delta}$ is the dual field strength tensor for $i=1,2$. 
Likewise, we can also obtain the chiral-odd or the helicity flip GPDs, which basically 
 uses the {gluon tensor operator} $\mathbf{S}F^{+i}(-z/2)F^{+j}(z/2)$ , as following:
\begin{align}
  \label{flip-gluon}
- \frac{1}{P^+} \int & \frac{d z^-}{2\pi}\,  e^{ix P^+ z^-}
  \langle p',\lambda'|\, {\mathbf S}
     F^{+i}(-{\textstyle\frac{z}{2}})\,
F^{+j}({\textstyle\frac{z}{2}})
  \,|p,\lambda \rangle \Big{|}_{\substack{z^+=0\\\mathbf{z}_{T}=0}} 
={\mathbf S}\,
\frac{1}{2 P^+}\, \frac{P^+ \Delta^j - \Delta^+ P^j}{2 M P^+}\nonumber \\
&
\times \bar{u}(p',\lambda') \Bigg[
  \Ta^g\, i \sigma^{+i}+\Tb^g\, \frac{P^+ \Delta^i - \Delta^+ P^i}{M^2} 
  +\Tc^g\, \frac{\gamma^+ \Delta^i - \Delta^+ \gamma^i}{2M} +
  \Td^g\, \frac{\gamma^+ P^i - P^+ \gamma^i}{M}\, 
   \Bigg] u(p,\lambda),
\hspace{2em} 
\end{align} 
 {where operator {\bf S} stands for the symmetrization in $i$ and $j$}. Also
{$u$ $\left(\bar{u}\right)$ are spinors for the spin-$\frac{1}{2} $ proton system along with $p$ $(p')$ and $\lambda$ $(\lambda')$  are their corresponding momenta and helicity of the initial (final) state of proton, respectively.  
$P^\mu=\frac{(p+p')^\mu}{2}$ is average momentum, $\Delta^\mu=p'^\mu-p^\mu$ is momentum transfer with $t=\Delta^2$ and $\xi=-\Delta^+/2P^+$ is the skewness parameter. }
\begin{figure}
	\centering
	\includegraphics[scale=0.23]{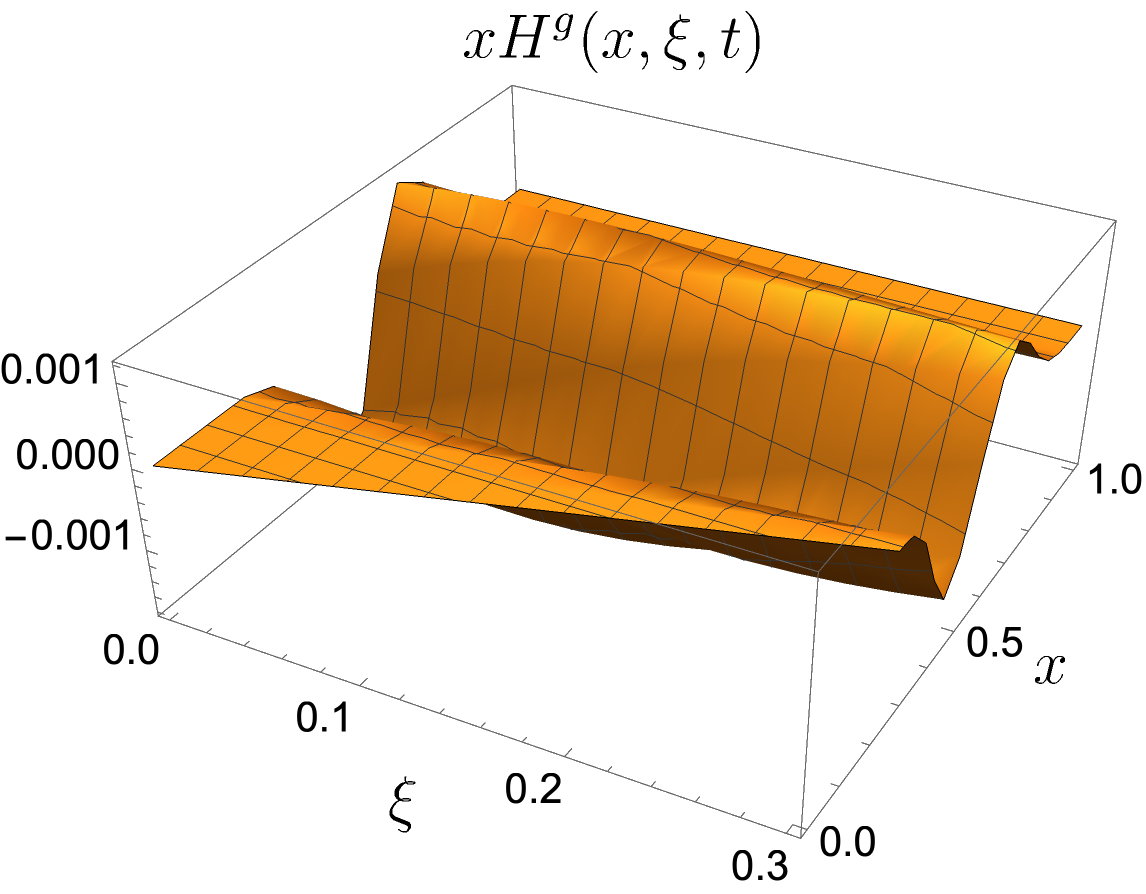}\hspace{0.2cm}
	\includegraphics[scale=0.23]{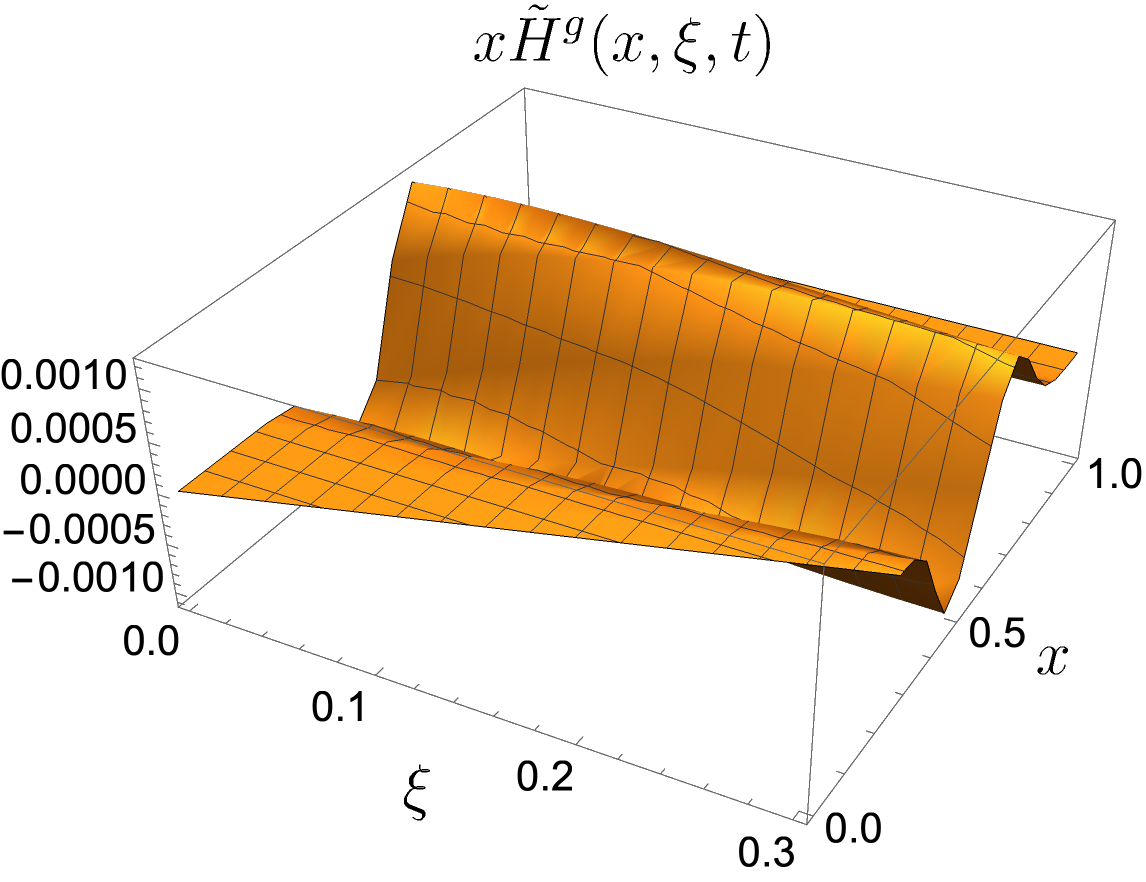}\hspace{0.2cm}
    \includegraphics[scale=0.23]{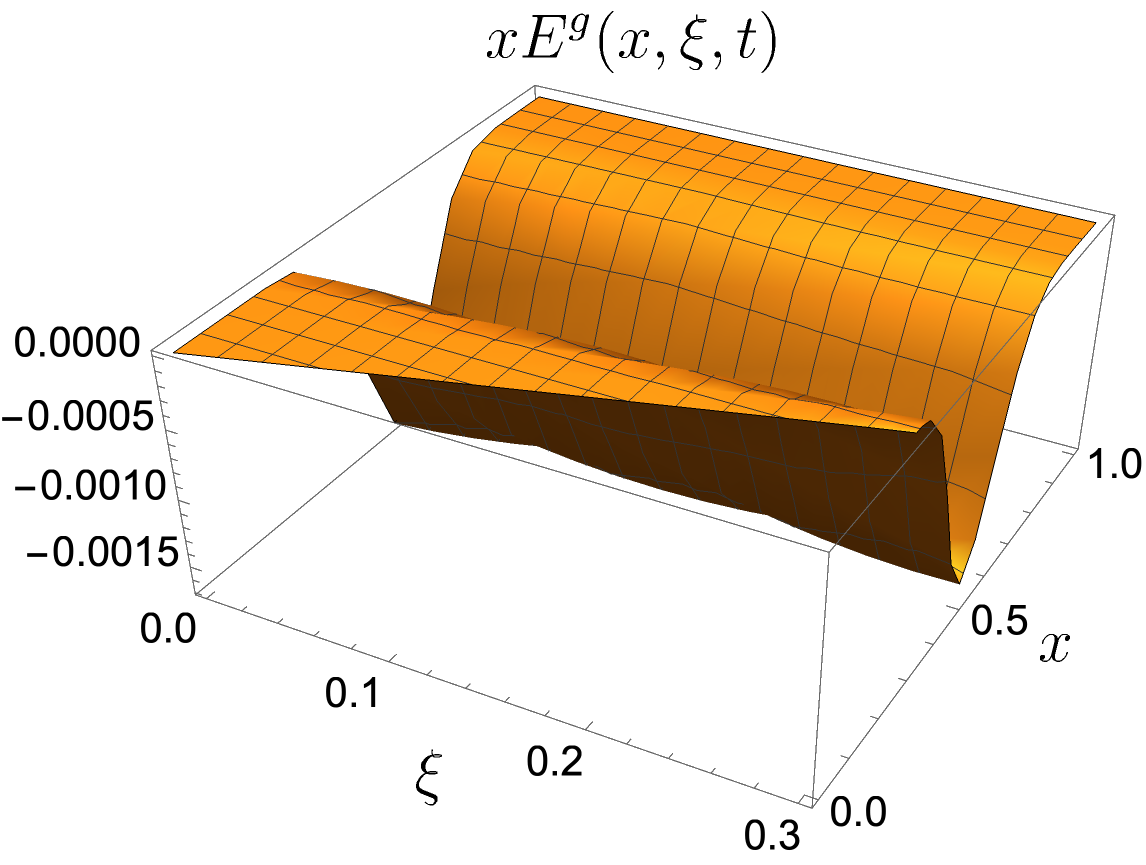}
    \vspace{0.2cm}
	\includegraphics[scale=0.23]{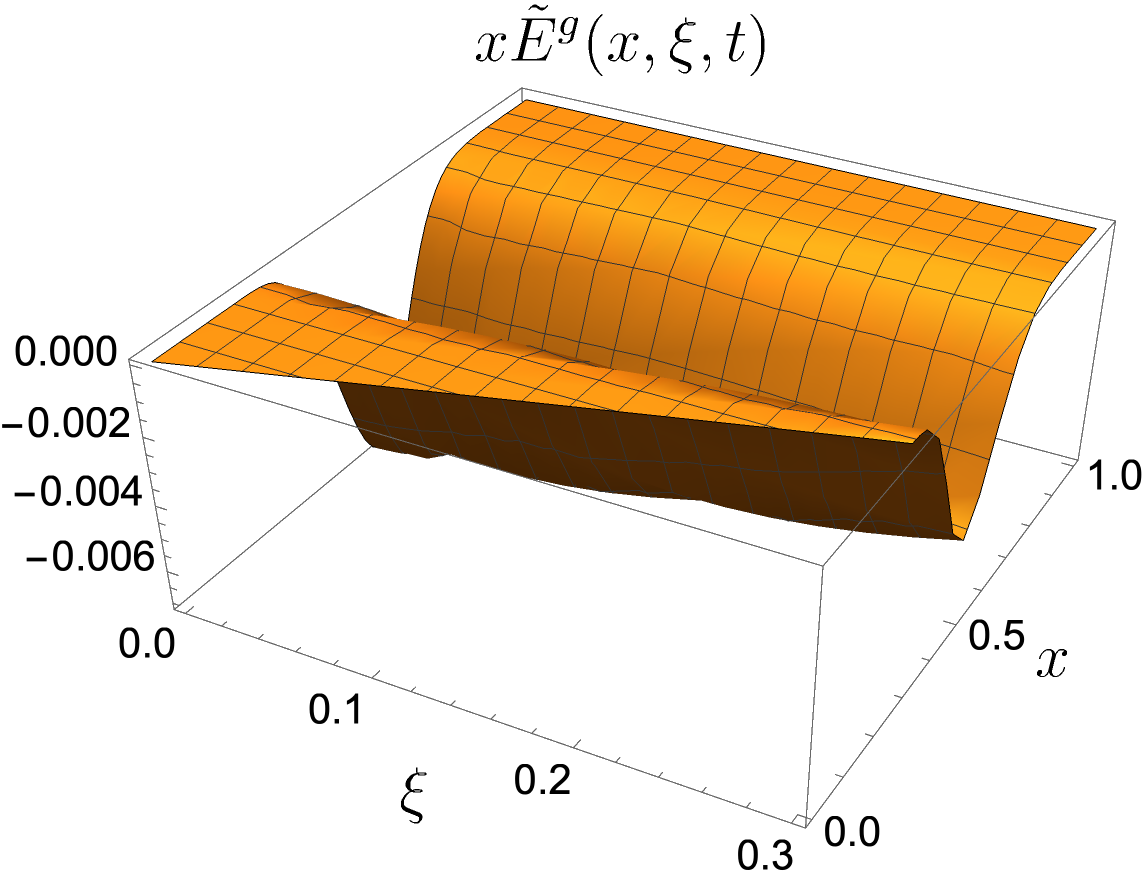}\hspace{0.2cm}
    \includegraphics[scale=0.23]{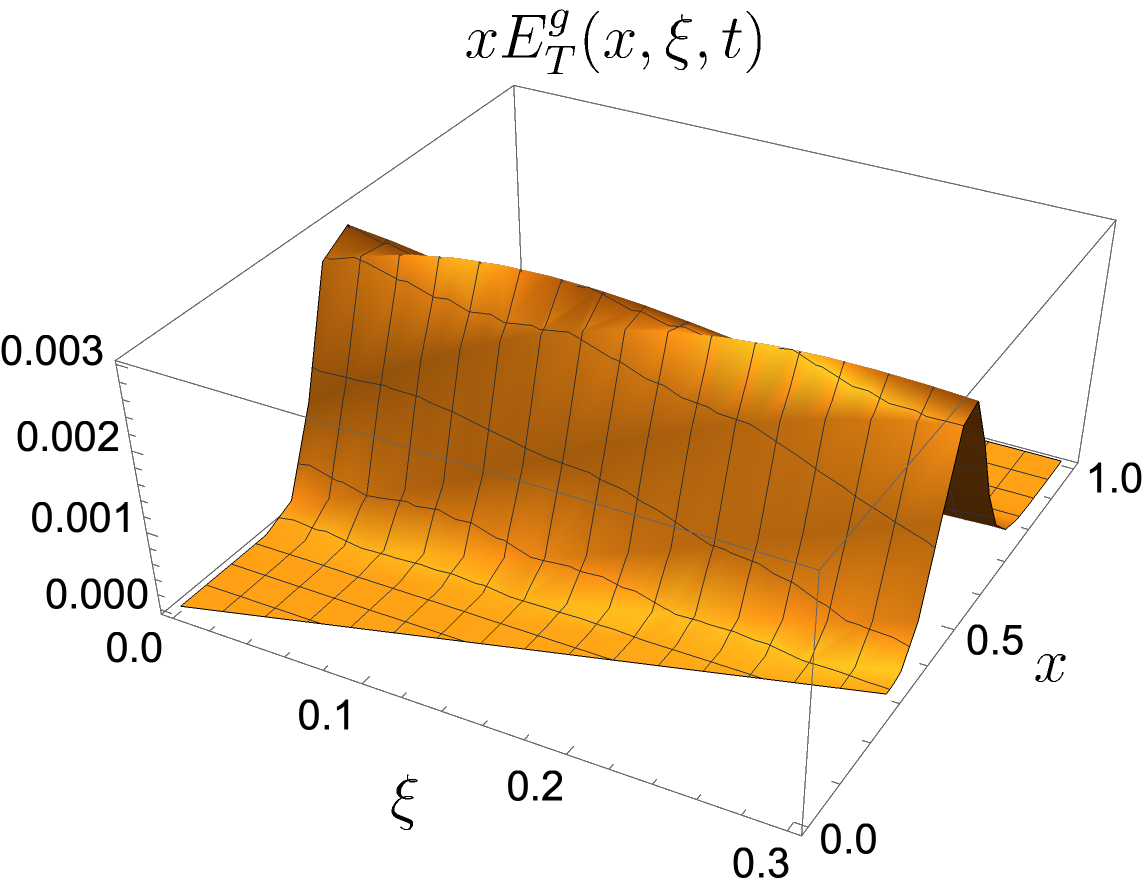} \hspace{0.2cm}
     \includegraphics[scale=0.23]{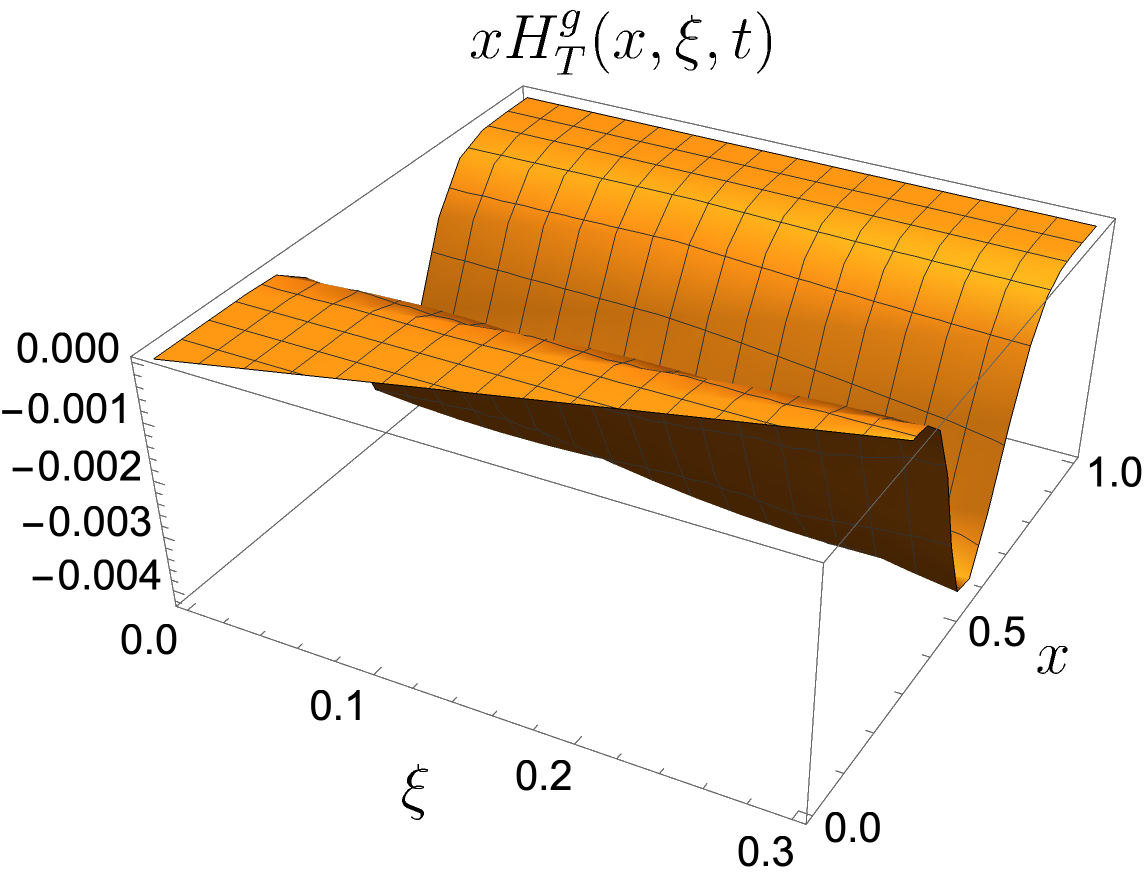} 
 \caption{For a fixed transverse momentum transfer $-\lvert t \rvert=3$ GeV$^2$, the model results for the gluon GPDs are displayed in three dimensions as functions of the gluon longitudinal momentum fraction $x$ and the skewness parameter $\xi$. The plots are taken from \cite{Chakrabarti:2024hwx}.}
 \label{fig:3Dnonzeroskewness}
\end{figure}
 In Fig. \ref{fig:3Dnonzeroskewness}, we present the results for the chiral even and odd GPDs for the range $0 < \xi < 0.3 $ and $\xi <x<1$ at the fixed transverse momentum transfer $-\lvert t \rvert=3$ GeV$^2$ in the DGLAP region. Our results and behaviour is consistent with \cite{Tan:2023kbl}. Using the GPDs information from the model, we also predicted the results for the gluon OAM. The total OAM $ J^{g}_{z}=\frac{1}{2}\int dx x\left[H^{g}(x,0,0)+E^{g}(x,0,0)\right]$ is found $J_{z}^{g} = 0.058$ in our model while in the BLFQ ~\cite{Lin:2023ezw} approach, it is reported $J^{g}_{z}|_{\text{BLFQ}} = 0.066$  at the scale $Q= 0.5$ GeV. We also evaluate the kinetic OAM $ L_{z}^{g}=\int dx \bigg{\{}\frac{1}{2} x\big[H^{g}(x,0,0)+E^{g}(x,0,0)\big]-\widetilde{H}^{g}(x,0,0)\bigg{\}}$ of the gluon which turned out to be negative i.e $L^{g}_{z}=-0.42$ in our model~\cite{Chakrabarti:2024hwx}, while another light-cone spectator model predicts $L^{g}_{z}=-0.123$~\cite{Tan:2023kbl}. 

\section{Conclusion}
In this contribution, we present the three-dimensional gluonic distribution within the proton, characterized using TMDs and GPDs within a light-front spectator model.
\section{Acknowledgment}
PC acknowledge the partial support from the conferences organizers and sponsors to attend and present the work at the conference. 



\end{document}